\documentclass[a4paper, 12pt]{article}
\author{Clovis Jacinto de Matos\footnote{ESA-HQ, European Space Agency, 8-10 rue Mario Nikis, 75015 Paris, France, e-mail: Clovis.de.Matos@esa.int}
}
\title{Gravitational force between two electrons in superconductors}

\begin{document}

\maketitle \begin{abstract} The attractive gravitational force
between two electrons in superconductors is deduced from the
Eddington-Dirac large number relation, together with Beck and
Mackey electromagnetic model of vacuum energy in superconductors.
This force is estimated to be weaker than the gravitational
attraction between two electrons in the vacuum.
\end{abstract}

\section{Introduction}

Suspicions about the underlying physical connections among the
parameters of the cosmos and particle physics arose as early as
1930 when Eddington \cite{Eddington} and Dirac \cite{Dirac}
investigated myriad occurrences of pure numbers of order
$10^{-40}$. These relations can be summarized in the following
expression \cite{Nottale2003}:
\begin{equation}
\alpha
\left(\frac{m_p}{m}\right)=\left(\frac{\Lambda^{-1/2}}{l_p}\right)^{1/3}.
\label{equ15}
\end{equation}
Where $l_p=(\frac{G \hbar}{c^3})^{1/2}$ is the Planck length,
$m_P=(\frac{\hbar c}{G})^{1/2}$ is the Planck mass, $m$ is the
electron mass, $\alpha=\frac{1}{4\pi\epsilon_0}\frac{e^2}{\hbar
c}$ is the fine structure constant,and $\Lambda$ is the
cosmological constant. Using the presently measured value of
$\Lambda_0=(1.29\pm 0.23)\times10^{-52}[m^{-2}]$
\cite{Spergel2003}, Equ.(\ref{equ15}) is valid within $1 \%$.

A modern formulation of the Eddington-Dirac large number
conjecture, equ.(\ref{equ15}), is possible in the framework of the
Planck-Einstein scale \cite{Christian, Clovis}, which corresponds
to typical energy scales for dark energy, and involves the
fundamental constants: $\Lambda, \hbar, c, k, G$. Explicitly one
has the following formulas at the Planck-Einstein scale:
\begin{equation}
E_{PE}=kT_{PE}=\Bigg(\frac{c^7\hbar^3 \Lambda}{G}\Bigg)^{1/4}=5.25
[meV]\label{e17}
\end{equation}
\begin{equation}
m_{PE}=\frac{E_{PE}}{c^2}=\Bigg(\frac{\hbar^3
\Lambda}{cG}\Bigg)^{1/4}=9.32\times10^{-39}[Kg]\label{e18}
\end{equation}
\begin{equation}
l_{PE}=\frac{\hbar}{M_{PE}c}=\Bigg(\frac{\hbar G}{c^3
\Lambda}\Bigg)^{1/4}=0.037[mm]\label{e19}
\end{equation}
\begin{equation}
t_{PE}=\frac {l_{PE}}{c}=\Bigg(\frac{\hbar
G}{c^7\Lambda}\Bigg)^{1/4}=1.26\times10^{-13}[s]\label{e20}
\end{equation}
\begin{equation}
\rho_{PE}=\frac{E_{PE}}{l_{PE}^3}=\frac{c^4 \Lambda}{G}=104
[eV/mm^3]
\end{equation}
The Planck temperature $T_P=\frac{1}{k}\sqrt{\frac{\hbar c^5}{G}}$
together with the Planck-Einstein
temperature,$T_{PE}=\frac{1}{k}\Bigg(\frac{c^7\hbar^3
\Lambda}{G}\Bigg)^{1/4}$, allow to rewrite equ.(\ref{equ15}) in
the following form:
\begin{equation}
\alpha\Bigg(\frac{m_P}{m}\Bigg)=\Bigg(\frac{T_P}{T_{PE}}\Bigg)^{2/3}.\label{ED}
\end{equation}
Thus it seems that the Eddington-Dirac large number relation takes
its "natural form" in the context of the Planck-Einstein scale.

The existence of dark energy in the universe, as indicated by
numerous astrophysical observations, represents one of the most
challenging problems in theoretical physics at present
\cite{Spergel2003,Peebles,Copeland,Padmanabhan}. A great variety
of different models exist for dark energy but none of these models
can be regarded as being entirely convincing so far. The
cosmological constant problem (i.e. the smallness of the
cosmologically observed vacuum energy density) remains an unsolved
problem. It is likely that the solution of this problem requires
new, so far unknown, physics.

Recent models of dark energy, such as the electromagnetic dark
energy model of Beck and Mackey \cite{Beck4}, produce potentially
measurable effects at laboratory scales, which are, however,
restricted to superconductors. In \cite{Beck4} a Ginzburg-Landau
theory is constructed that generates a cutoff for the
gravitational activity of vacuum fluctuations. Generally it is
assumed in this model that vacuum fluctuations of any particle can
exist in two different phases: A gravitational active one
(contributing to the cosmological constant $\Lambda$) and a
gravitationally inactive one (not contributing to $\Lambda$). The
model exhibits a phase transition at a critical frequency which
makes the dark energy density in the universe small and finite.
The above approach has many analogies with the physics of
superconductors, and in particular it allows for a possible
interaction between dark energy and Cooper pairs. Using the vacuum
energy predicted by this model in superconducting materials
together with the Eddington-dirac conjectured relationship
equ.(\ref{equ15}), we estimate the gravitational force between two
electrons in superconducting materials, and compare it with its
value in vacuum.

\section{Electron's gravito-electromagnetic coupling from the Eddington-Dirac large number relation}

From Eddington-Dirac large number relation, Equ.(\ref{equ15}), we
deduce the coupling between the gravitational and the
electromagnetic interaction between two electrons in vacuum, in
function of the electron Compton wavelength
$\lambda_c=\frac{\hbar}{mc}$, the cosmological constant, and the
square of the fine structure constant, $\alpha$

\begin{equation}
\frac{\alpha_g}{\alpha}=\alpha^2 \lambda_c \Lambda^{1/2}.
\label{equ18}
\end{equation}

Where $\alpha_g= \frac{G m^2} {\hbar c}=(\frac{m}{m_P})^2$ is the
electron's gravitational fine structure constant. Note that a null
value of the cosmological constant would imply that electrons
could note simultaneously generate gravitational and
electromagnetic fields.

A non-vanishing cosmological constant (CC) $\Lambda$ can be
interpreted in terms of a non-vanishing vacuum energy density
\begin{equation}
\rho_{vac\Lambda}=\frac{c^4}{8\pi G} \Lambda ,\label{e14}
\end{equation}
which corresponds to dark energy. The small astronomically
observed value of the CC, $\Lambda=1.29\times10^{-52}[1/m^2]$
\cite{Spergel2003}, and its origin remain a deep mystery. This is
often call the CC problem, since with a cutoff at the Planck scale
the vacuum energy density expected from quantum field theories
should be larger by a factor of the order $10^{120}$, in complete
contradiction with the observed value. Substituting equ(\ref{e14})
into equ.(\ref{equ18}), it is possible to express the coupling
between gravitational and electromagnetic forces between two
electrons in vacuum in function of the dark energy density.
\begin{equation}
\frac{\alpha_g}{\alpha}=\frac{2(2\pi
G)^{1/2}}{c^2}\alpha^2\lambda_c\rho_{vac\Lambda}^{1/2}\label{equat1}
\end{equation}

\section{Electromagnetic model of vacuum energy in superconductors}

To solve the CC problem, in \cite{Beck4} a model of dark energy
was suggested that is based on electromagnetic vacuum fluctuations
creating a small amount of vacuum energy density. One assumes that
photons (or any other bosons), with zeropoint energy
$\epsilon=\frac{1}{2} h \nu$, can exist in two different phases: A
gravitationally active phase where the zeropoint fluctuations
contribute to the cosmological constant $\Lambda$, and a
gravitationally inactive phase where they do not contribute to
$\Lambda$ \cite{Beck4, Beck,Beck2,Beck3}. This is described in
\cite{Beck4} by a Ginzburg-Landau type of theory. As shown in
\cite{Beck4}, this type of model of dark energy can lead to
measurable effects in superconductors, via a possible interaction
with the Cooper pairs in the superconductor.

Here we introduce an additional hypotheses with respect to the
original Beck and Mackey model: \begin{enumerate} \item {the
vacuum energy density contained in superconductors can be
different from the energy density observed in the universe.}
\end{enumerate}

Beck and Mackey's Ginzburg-Landau-like theory leads to a finite
dark energy density dependent on the frequency cutoff $\nu_c$ of
vacuum fluctuations:
\begin{equation}
\rho^*=\frac{1}{2}\frac{\pi h}{c^3}\nu_c^4\label{e15}
\end{equation}
In vacuum one may put $\rho^*=\rho_{vac\Lambda}$, from which the
cosmological cutoff frequency $\nu_{cc}$ is estimated as
\begin{equation}
\nu_{cc}\simeq2.01 THz\label{e16}
\end{equation}
The corresponding "cosmological" quantum of energy is:
\begin{equation}
\epsilon_{cc}=h\nu_{cc}=8.32 meV\label{e166}
\end{equation}
In the interior of superconductors, according to assumption 1.
above, the effective cutoff frequency can be different. This is
due to interaction effects between the two Ginzburg-Landau
potentials (that of the superconducting electrons and that of the
dark energy model) \cite{Beck4}. The effect can be seen in analogy
to polarization effects of ordinary electromagnetic fields in
matter: In matter the electric field energy density is different
as compared to the vacuum. Similarly, in superconductors the
effectice dark energy density (represented by gravitationally
active zeropoint fluctuations) can be different as compared to the
vacuum. Our model allows for the gravitational analogue of
polarization.

An experimental effort is currently taking place at University
College London and the University of Cambridge to measure the
cosmological cutoff frequency through the measurement of the
spectral density of the noise current in resistively shunted
Josephson junctions, extending earlier measurements of Koch et al.
\cite{Koch}.

In \cite{Beck4} the formal attribution of a temperature $T$ to the
graviphotons is done by comparing their zeropoint energy with the
energy of ordinary photons in a bath at temperature $T$:
\begin{equation}
\frac{1}{2} h\nu=\frac{h\nu}{e^{\frac{h\nu}{kT}}-1}\label{e16a}
\end{equation}
This condition is equivalent to
\begin{equation}
h\nu=\ln3kT\label{e17}
\end{equation}
Substituting the critical transition temperature $T_c$ specific to
a given superconductive material into Eq.(\ref{e17}), we can
calculate the critical frequency characteristic for this material:
\begin{equation}
\nu_c=\ln3 \frac{kT_c}{h}\label{e18}
\end{equation}
For example, for Niobium with $T_c=9.25$K we get $\nu_c=0.212$
THz. If we use the cosmological cutoff frequency, equ.(\ref{e16}),
in Eq.(\ref{e18}) we find the cosmological critical temperature
$T_{cc}$:
\begin{equation}
T_{cc}=87.49K\label{e19}
\end{equation}
This temperature is characteristic of the BSCCO High-$T_c$
superconductor.

Substituting equ.(\ref{e18}) into equ.(\ref{e15}), the vacuum
energy stored in a given superconductor is obtained from a
Stephan-Boltzmann type law, showing a dependence on the fourth
power of the superconductor's critical transition temperature:
\begin{equation}
\rho^*=\frac{\pi (\ln 3)^4}{2}\frac{k^4}{(ch)^3}T_c^4\label{e12}
\end{equation}

\section{The gravitational force law between two electrons in superconductors}

Using Beck and Mackey's electromagnetic vacuum energy density in
superconductors $\rho^*$, equ.(\ref{e12}), instead of the
cosmological energy density $\rho_{vac\Lambda}$, equ.(\ref{e14}),
in the expression defining the coupling between gravitational and
electromagnetic forces between two electrons in vacuum,
equ.(\ref{equat1}), we obtain this coupling in a superconducting
environment:
\begin{equation}
\frac{\alpha_g}{\alpha}=\alpha^2\frac{(ln 3)^2}{(2
\pi)^{1/2}}\frac{k^2}{m m_P c^4}T_c^2\label{equat2}
\end{equation}
Taking the geometric mean value between the Planck scale and the
electron scale, it is possible to define an intermediate energy
scale at $7.9 [GeV]$:
\begin{equation}
m_i=(m m_P)^{1/2}\label{equat3}
\end{equation}
substituting equ.(\ref{equat3}) into equ.(\ref{equat2}), we
obtain:
\begin{equation}
\frac{\alpha_g}{\alpha}=\alpha^2\frac{(ln 3)^2}{(2
\pi)^{1/2}}\Bigg( \frac {T_c}{T_i}\Bigg)^2\label{equat4}
\end{equation}
where $T_i=m_i c^2/k\sim9\times 10^{20}[K]$ is the temperature
associated with the intermediate energy scale defined from
equ.(\ref{equat3})

Equ.(\ref{equat4}) indicates that the coupling between
gravitational forces and electromagnetic forces between two
electrons in superconductors is weaker than  in vacuum, recovering
its classical value at the temperature, $T_c\sim87.5 [K]$ (BSCCO's
critical transition temperature).

Assuming that the electromagnetic fine structure constant does not
change, with respect to its classical value. We can only explain a
deviation from the classical gravito-electromagnetic coupling in
terms of a suitable scale transformation: The gravitational
constant $G=\hbar c/m_P^2$ formally becomes much weaker in a
superconductor than in vacuum if $m_P$ is replaced by a higher
value $m_{P eff}$. Thus from equ.(\ref{equat4}) we deduce the
effective gravitational constant between two electrons in a given
superconducting material:
\begin{equation}
G_{eff}=\frac{\hbar c}{m_{P
eff}^2}=\frac{\alpha^2}{4\pi\epsilon_0}\frac{(ln 3)^2}{(2
\pi)^{1/2}} \Bigg(\frac{e}{m}\Bigg)^2 \Bigg( \frac
{T_c}{T_i}\Bigg)^2\label{equat6}
\end{equation}

Substituting this expression in Newton's law of gravitation, and
applying it to a system of two electrons separated by a distance
$r$ inside a superconductor. We conclude that their mutual
gravitational attraction force, $F_{SC}$, will be smaller than the
gravitational force, $F_0$, that the same electrons, separated by
the same distance, will exert on each other when they are in
vacuum and the classical value of the gravitational constant
$G_0=6.67\times 10^{-11} [Nm^2/Kg^2]$ holds:
\begin{equation}
\frac{F_{SC}}{F_0}=\frac{G_{eff}}{G_0}=\frac{\alpha^2}{4\pi\epsilon_0
G_0}\frac{(ln 3)^2}{(2 \pi)^{1/2}} \Bigg(\frac{e}{m}\Bigg)^2
\Bigg( \frac {T_c}{T_i}\Bigg)^2\label{equat7}
\end{equation}

For the case of two electrons located in superconducting Niobium,
for which $T_c=9.25 [K]$, the numerical estimation of
equ.(\ref{equat7}) is:
\begin{equation}
\frac{F_{SC}}{F_0}\sim10^{-2}\label{equat8}
\end{equation}
This prediction is difficult to test experimentally, since the
electron's mass only contributes marginally to the overall mass of
atoms and of macroscopic bodies: ($m_{electron}/m_{proton}=
5.4\times 10^{-4}$).

At this point one remark is in order: Our theoretical derivation
presented in this paper strictly speaking holds only for
electrons, because equ.(\ref{equ15}) is only valid for electrons
(in that expression $m$ must always be the electron mass).
Therefore equ.(\ref{equat7}) should not apply to the other
particles (protons, atoms) the superconductor consists of. This
ultimately could be understood as being a direct consequence of
the spontaneous breaking of the principle of general covariance in
superconductors \cite{pos}.

\section{conclusion}

From the Eddington-Dirac Large number conjecture,
equ.(\ref{equ15}), together with Beck and Mackey electromagnetic
model for dark energy density in superconductors, equ.(\ref{e12}),
it is shown that the gravitational force between two electrons in
superconductors should deviate from the classical law of
gravitational attraction between two electrons in vacuum,
according to equ.(\ref{equat7}). This effect is estimated
difficult to be detected within current experimental capabilities.

\section{Acknowledgment}

The author is grateful to Prof. Christian Beck for useful comments
on the physical interpretation of the proposed conjecture.

\end{document}